\documentclass[twocolumn,showpacs,preprintnumbers,amsmath,amssymb]{revtex4}
\usepackage{graphicx}% Include figure files
\usepackage{dcolumn}% Align table columns on decimal point
\usepackage{bm}% bold math
\begin{document}

\preprint{APS/123-QED}

\title{Universal scaling and quantum critical behavior of CeRhSb$_{1-x}$Sn$_{x}$}

\author{Andrzej \'Slebarski}
\email{slebar@us.edu.pl}
\affiliation{Institute of Physics, University of Silesia, Uniwersytecka 4, 40-007 Katowice, Poland}

\author{Jozef Spa\l{}ek}
\email{ufspalek@if.uj.edu.pl}
\affiliation{Marian Smoluchowski Institute of Physics, Jagiellonian University, Reymonta 4, 30-059 Krak\'ow, Poland}

\date{\today}% It is always \today, today,

\begin{abstract}
We propose a universal scaling $\rho \chi=const$ of the electrical resistivity $\rho$ with the inverse magnetic 
susceptibility $\chi^{-1}$ below the temperature of the {\em quantum-coherence onset\/} for the Ce
4f states in CeRhSb$_{1-x}$Sn$_{x}$. In the regime, where the Kondo gap disappears 
$(x\simeq 0.12)$, the system forms a {\em non-Fermi liquid\/} (NFL), which transforms into
a Fermi liquid at higher temperature. The NFL behavior is attributed to the presence of a novel {\em quantum critical point\/} (QCP)
at the Kondo insulator - correlated metal boundary. The divergent behavior
of the resistivity, the susceptibility, and the specific heat has been determined when approaching QCP from the metallic side.
\end{abstract}

\pacs{71.27.+a, 72.15.Qm, 71.30+h}% PACS, the Physics and Astronomy
\maketitle

The compounds CeRhSb and CeRhSn, both with strongly correlated nature of 4f electrons, form 
unprecedented types of ground state: CeRhSb represents the {\em Kondo semiconductor\/} with 
a gap $\Delta\simeq 7$~K [1] and CeRhSn is a {\em non-Fermi\/} (non-Landau) quantum liquid [2].

The fundamental question is whether by studying CeRhSb$_{1-x}$Sn$_{x}$ the two states can be
related through a corresponding {\em quantum critical
point\/} (QCP) or a phase boundary located in between. In that respect, the 
quantum critical points between 
antiferromagnetic metal and either superconducting \cite{3} or paramagnetic [4] 
heavy-fermion states have been successfully identified
play an important role in the discussion of the origin of exotic superconductivity [3]. Also, the
existence of the QCP between the Kondo insulator and an itinerant antiferromagnet has been suggested
[5] as a function of either the Kondo coupling or pressure. 

In this paper
we show directly that a transition of this type exhibits indeed a well defined quantum critical behavior as a function of 
the number of carriers. The number of the carriers per $4f$ electron is a crucial factor when examining the formation 
of the collective Kondo-singlet state (cf. famous "exhaustion" theorem of Nozieres [6]).
To address these basic topics, we have studied the series CeRhSb$_{1-x}$Sn$_{x}$, in which the number 
of valence electrons diminishes by one per formula when substituting Sn atom for Sb. The characterization 
of such QCP allows also to compare the insulator-metal transition for the 
{\em Mott-Hubbard\/} systems and that for the {\em Anderson-Kondo\/} lattices. We show that the 
behavior in Ce compounds differs from that for the charge-transfer Mott-Hubbard systems [6], even 
though the hybridization between the strongly correlated and valence states is common to both classes. It 
implies that the {\em Mott-Hubbard\/} and {\em Anderson-Kondo} lattice systems may belong to separate
{\em universality classes\/} of systems with quantum phase transitions.

\begin{figure}
\includegraphics[width=0.35\textwidth]{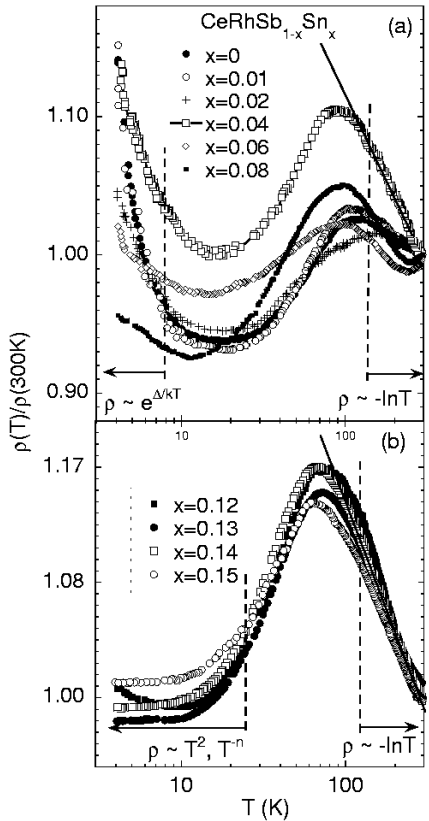}
%{f1.eps}% Here is how to import EPS art
\caption{\label{fig:r1} Temperature dependence of the relative resistivity $\rho$ for CeRhSb$_{1-x}$Sn$_{x}$
systems for $x\leq 0.08$ (a) and $0.12\leq x\leq 0.15$ (b). The regimes, where $\rho(T)\sim\exp 
(\Delta/k_{B}T),\,\sim-\ln T$, and $\sim T^{2}$ are marked by the corresponding solid lines.}
\end{figure}

We first demonstrate, on the basis of 
experimental-data correlation, a novel type 
of scaling between the electrical resistivity $\rho$ and the magnetic susceptibility $\chi$ in the 
{\em quantum-coherence regime\/} for the Kondo semiconductors CeRhSb$_{1-x}$Sn$_{x}$, with 
$x\leq 0.12$. The reference resistivity data are displayed in Fig. 1ab. We are interested in the 
properties below the temperature $T_{max}$, where the resistivity has a maximum. For $T>T_{max}$ 
a pronounced $\ln T$ behavior of $\rho$ should be noted for all the samples studied [7]. However, for 
$T\ll T_{max}$ a pronounced exponential increase of $\rho\sim\exp (\Delta/T)$, representing
a nondegenerate semiconductor, can be fitted well for $x<0.12$; the corresponding gap values
$\Delta=\Delta(x)$ are listed in Table I. The properties of the metallic state for $x>0.12$ will be discussed 
below.

\begin{table}
\caption{\label{tab:table1}The Kondo gap $\Delta$ for the CeRhSb$_{1-x}$Sn$_{x}$ samples for $x\leq 0.12$.}
\begin{tabular}{|c|c|}
\hline
\mbox{\phantom{aaaaa}}x\mbox{\phantom{aaaaa}} & $\Delta(K)$ \\
\hline
0 & 6.70 \\
0.01 & 1.80 \\
0.02 & 0.64 \\
0.04 & 0.70 \\
0.06 & 0.30 \\
0.08 & 0.22 \\
0.10 & 0.10 \\
0.12 & 0.10 \\
\hline
\end{tabular}
\end{table}

In Fig. 2 we show exemplary data (for $x=0$) of the scaling of $1/\rho$ with $\chi$; the inset provides the 
detailed behavior. One should note that in order to obtain such a good scaling, we have to subtract from the 
measured $\chi$ the impurity Curie-law contribution $(nC/T)$ with $n$ 
in the regime $0.004\div 0.008$ (depending on the sample), which represents a standard 
procedure in these and related compounds [8]. Also, then $\chi\rightarrow 0$ with $T\rightarrow0$,
demonstrating that the Kondo semiconductors can be regarded at $T=0$ as either a quasiparticle band 
insulators (with a band gap renormalized by the electron correlations [9])
or as a collective Kondo-lattice insulator [10] with a singlet spin state in the ground state. The vanishing $\chi$
in the ground state distinguishes the Kondo insulators from the Mott-Hubbard insulators,
for which the correlated electrons have unpaired spins and thus order magnetically when $\Delta>0$.

\begin{figure}
\includegraphics[width=0.5\textwidth]{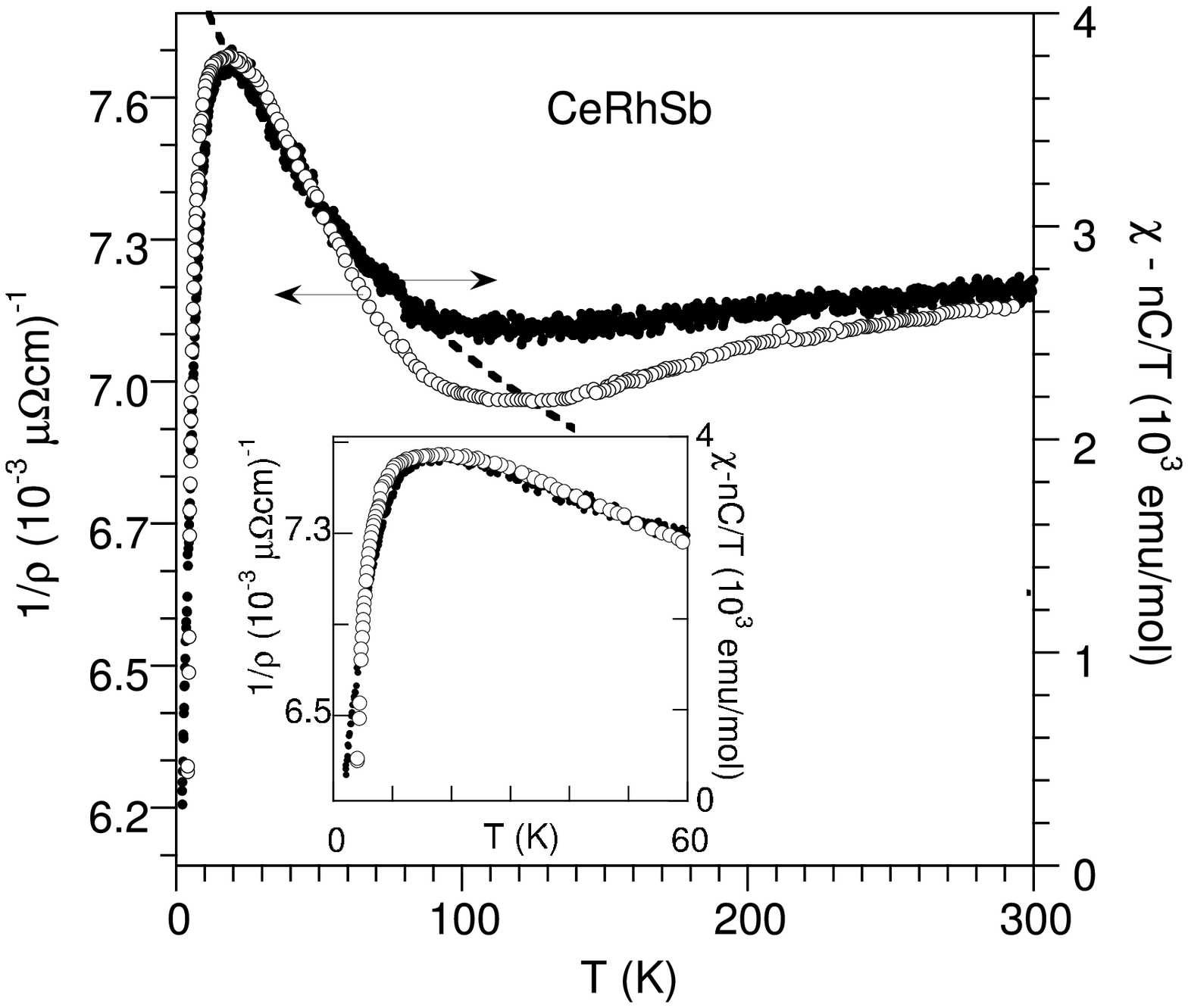}% Here is how to import EPS art
\caption{\label{fig:r2} Temperature dependences of the inverse resistivity $\rho^{-1}$ (left scale) and of the 
paramagnetic susceptibility $\chi$ (with the impurity-contribution $nC_{0}/T$, $n=0.004$, 
subtracted) for CeRhSb. The dashed line displays the Curie-Weiss contribution to $\chi$.
The inset: $\rho^{-1}$ and $\chi$ on an expanded scale.}
\end{figure}

The details of the scaling $\rho\chi=const(x)$ are demonstrated explicitly in Fig. 3, where the linear 
dependence between $\chi$ and $\rho$ is shown (note that we have coalessed the data to a common origin in each case). The inset
illustrates the value of the constant $R_{S}=\rho\chi$ as a function of $x$. In the regime $x\lesssim 0.06$
it has a universal value $R_{S}\simeq 0.1\,\mu\Omega$~cm~emu/mol. The existence of the scaling
displayed in Fig. 3 demonstrates that all the systems exhibit a common universal property, i.e. can be regarded
as belonging to a single {\em class\/}. It remains to be discussed next whether they form a {\em phase\/} 
in the thermodynamic sense. This can be done only by determining the nature of the boundary separating it from 
the metallic phase setting in for $x>0.12$.

\begin{figure}
\includegraphics[width=0.45\textwidth]{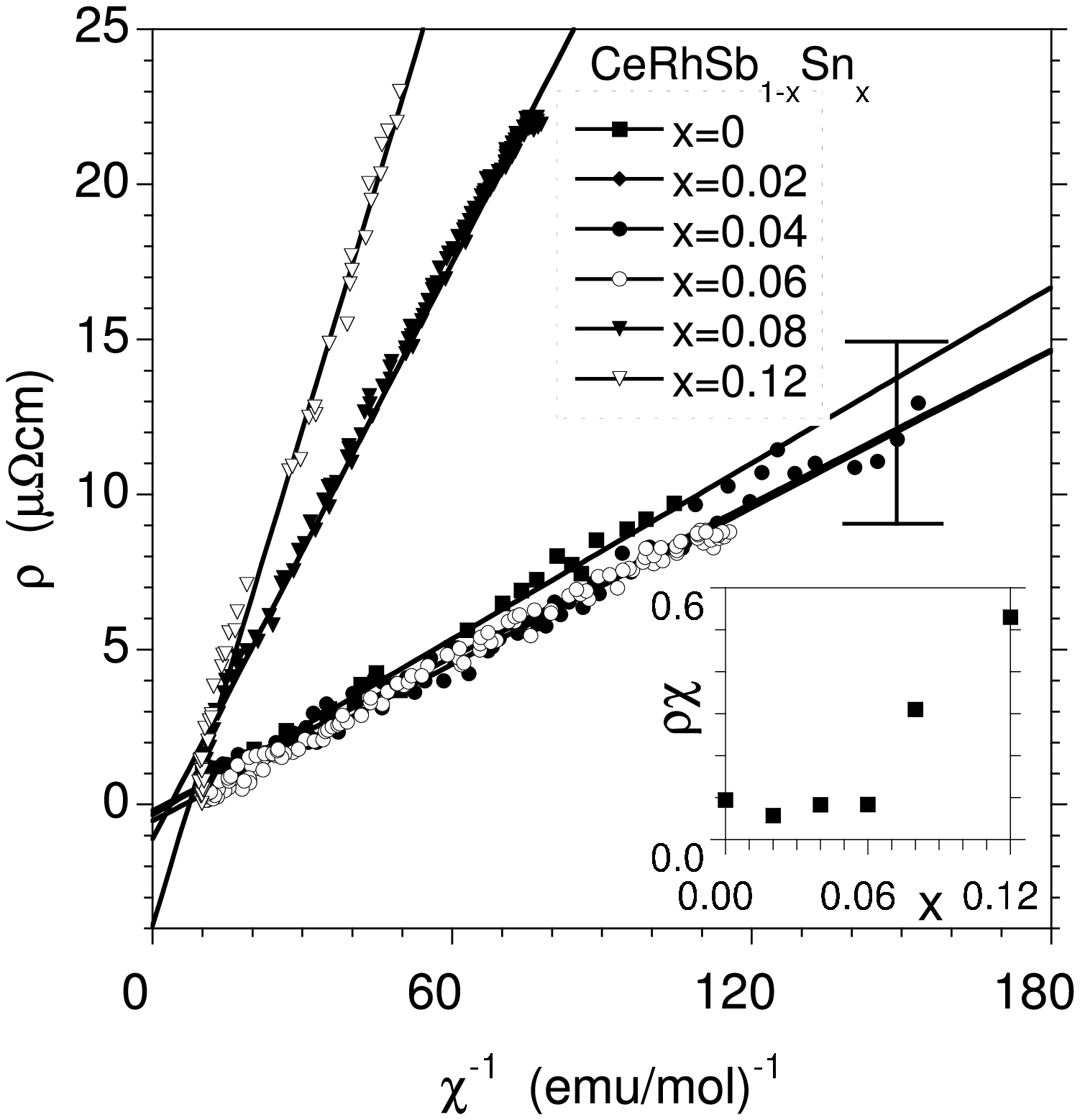}% Here is how to import EPS art
\caption{\label{fig:r3} Linear scaling law between the resistivity $\rho$ and the susceptibility $\chi$
for the systems with nonvanishing Kondo gap. The quantities label the corresponding $\rho$
and $\chi$ values shifted to the zero value at the minimal positions. The error bars represent the
inaccuracy of $\rho$ measurements.}
\end{figure}

To characterize the metallic phase we have plotted in Fig. 4 the resistivity data as a function of $T^{2}$.
This dependence is well fulfilled for the higher temperatures $T>10$~K. In the $T$ range below 
about 10~K (see the inset), a clear deviation from the Landau-Pomeranchuk-Baber law $\rho=\rho_{0}+AT^{2}$
for the Fermi liquid can be observed, particularly for the concentration $x\simeq 0.13$, where the gap has barely 
disappeared. The fitting to the $\sim T^{-n}$ dependence, with $n\simeq 0.1$, should be regarded only as an indicative factor 
of an incipient {\em non-Fermi-liquid (NFL) behavior\/}, requiring an additional test. Alternatively, these low-$T$ data can be 
represented by the $\ln T$
dependence.

\begin{figure}
\includegraphics[width=0.5\textwidth]{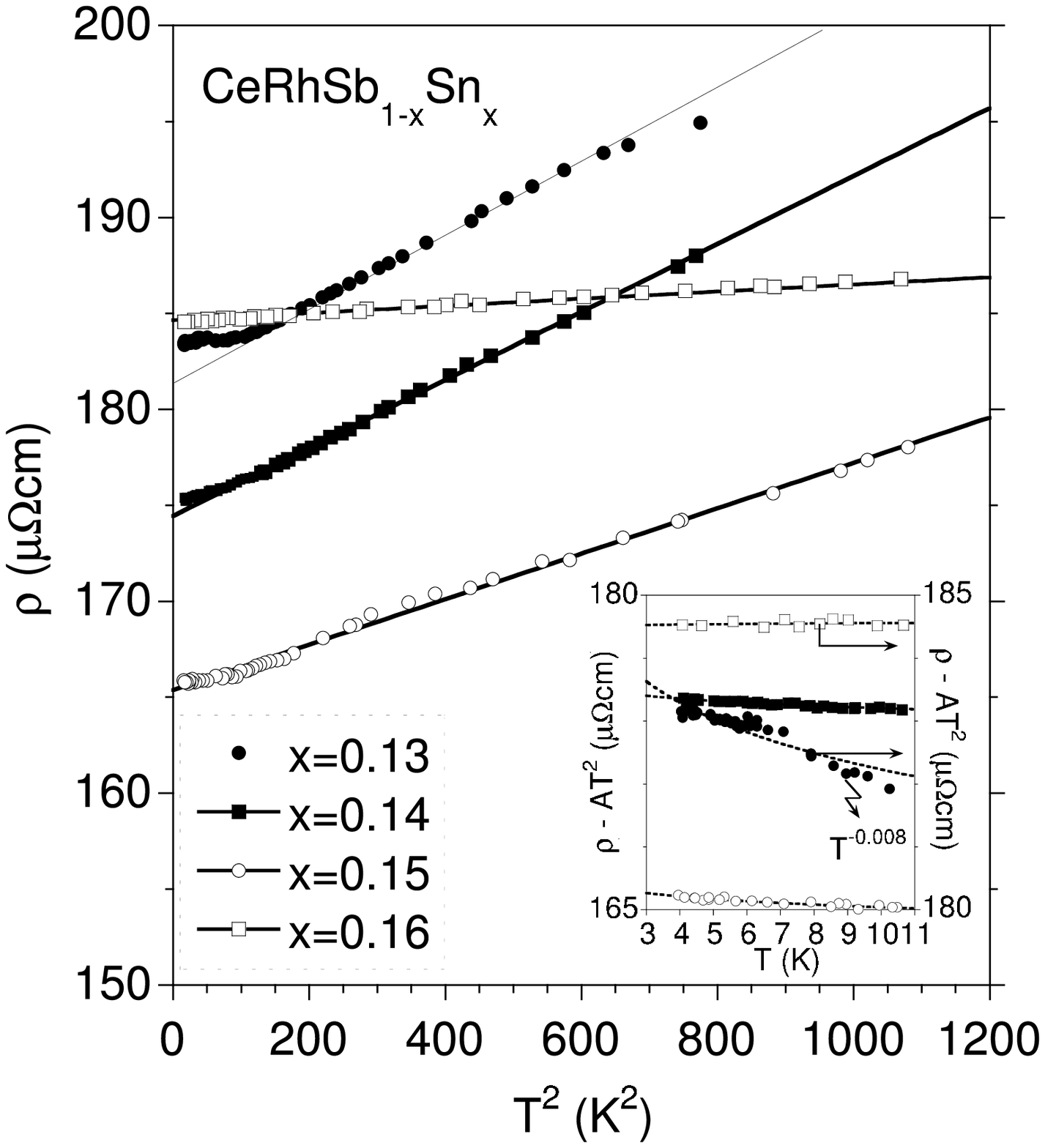}% Here is how to import EPS art
\caption{\label{fig:r4} Resistivity vs. $T^{2}$ for the samples with $0.13\leq x\leq 0.16$. The inset: 
non-Fermi-liquid-type scaling $T^{-n}$, $n=0$, for the sample $x=0.13$, i.e. close to the critical
point where the Kondo gap disappears. The lines are the fitted curves.}
\end{figure}

To determine thermodynamic properties of the non-Fermi liquid we have to discuss the temperature dependence of
equilibrium characteristics in the vicinity of the critical point, in our situation placed at $x\simeq 0.12$
and $T=0$. For that purpose, we have shown in Fig. 5 the temperature dependence of $\chi(T)$ for
$0.13\leq x\leq 0.16$. In each case, a clear dependence $\chi=aT^{-m}$ is observed. The values of the
parameters are listed in Table II. The quantities $\chi^{-1}$ and $\rho$ now do not
obey the linear type of relation shown in Fig. 3 for KI state. Also, now $\chi$ diverges, whereas $\rho$ approaches the finite value 
$\rho_{0}$ when $T\rightarrow 0$. The divergence of $\chi$ signals also a magnetic phase transition,
probably to weakly ferromagnetic or Griffiths phases [11]. The onset of magnetism signals a spontaneous
symmetry breakdown at $T=0$ needed for the {\em critical point\/} to be well defined. The onset of magnetism 
cannot be related to the Kondo insulating phase, since then $\chi\rightarrow 0$. So, whereas in Mott-Hubbard 
systems the insulating phase is magnetic (very often antiferromagnetic) and the metallic phase 
can be paramagnetic, here the Kondo insulator is nonmagnetic, whereas the metallic phase 
is magnetic, most probably with a very small moment, as in many heavy-fermion systems.
The existence of a weakly ferromagnetic phase in Sn-rich samples has been 
observed earlier [2], so we will not discuss it in detail here.

We have also performed the specific heat $(C_{p})$ measurements on the sample with $x=0.13$. The data in the temperature range
$10-20$~K can be fitted to the dependence $C_{p}/T=\gamma+\beta T^{2}$, with $\gamma=63.3$~mJ/K$^{2}$mol.
Below $10$~K the data start deviating from this dependence and the difference can be fitted to the formula
$C_{p}/T=bT^{-s}$, with $b=143.6$~mJ/K$^{1.7}$, and $s\simeq 0.3$.
So again, as in the case of the resistivity, the NFL critical regime appears for $T\rightarrow 0$ from the gross
Fermi-liquid-like state. 

One additional feature of the analysis should mentioned. The $\chi(T)$ data for $T> 15$~K and in the metallic phase 
($x\geq 0.13$) can be parametrized in the form: $\chi(T)=\chi_{0}+C/(T-\Theta)$, where value of the 
$\chi_{0}\simeq 1.7\div 3.5\cdot 10^{-3}$~emu/mol, a value in the range for the heavy fermions (cf. Table II).

\begin{table*}
\caption{\label{tab:table2}Susceptibility and the resistivity parametrizations of the samples in the {\em metallic\/}
regime $(x\geq 0.13)$.}
\begin{tabular}{|c|c|c|c|c|}
\hline
\mbox{\phantom{aa}}x\mbox{\phantom{aa}} & $\chi =\chi_{0}+C/(T-\Theta)$ & $\chi = aT^{-m}$ & $\rho = \rho_{0}+AT^{2}$ & $\rho-AT^{2}\sim T^{-n}$ \\
\cline{2-5}
 &  $\chi_{0}\,\left(10^{3}\,\frac{\mbox{emu}}{\mbox{mol}}\right)\;\;\mbox{\phantom{aa}}\Theta(K)\;\;\mbox{\phantom{aaaaa}}C \left(
 \frac{\mbox{emu K}}{\mbox{mol}}\right)$
 & $10^{3}\,a\;\;\;\;m$ & 
 $A(10^{2}\,\mu\,\Omega\,\mbox{cm}\,K^{2})$ & $10^{3}\,n$ \\
 \hline
0.13 & $\mbox{\phantom{aaaa}}2.6\;\;\;\mbox{\phantom{aaaaaa}} -7.0\;\;\;\mbox{\phantom{aaaaaaaa}}  0.043$ & $6.0\;\;\;\;\mbox{\phantom{aa}} 0.17$ & 1.78 & 8.0 \\
0.14 & $\mbox{\phantom{aaaa}}1.7\;\;\;\mbox{\phantom{aaaaaa}} -9.1\;\;\;\mbox{\phantom{aaaaaaaa}} 0.021$ & $4.3 \;\;\;\;\mbox{\phantom{aa}} 0.17$ & 1.69 & 2.8 \\
0.16 & $\mbox{\phantom{aaaa}}3.5\;\;\;\mbox{\phantom{aaaaaa}} -7.6\;\;\;\mbox{\phantom{aaaaaaaa}} 0.038$ & $6.0 \;\;\;\; \mbox{\phantom{aa}}0.09$ & 0.186 & 1.3 \\
\hline
 & 15 K $\leq$ T $\leq$ 90 K & T$<$ 10 K & 30 K $\geq$ T $\geq$ 10 K &  T $<$ 10 K \\
\hline
\end{tabular}
\end{table*}

We now discuss the results in a broader scope. First, the activated 
behavior of the electronic conductivity comes from the circumstance that the carrier concentration $n_{c}$
for $x\leq0.12$ increases by thermal excitation over the Kondo gap $\Delta$.
The dependence $\rho\sim \exp (\Delta/T)$ is obeyed even for temperature sizably higher than $\Delta$
and is in agreement with the well known fact that the Kondo semiconductors represent the systems with a low carrier concentration.
Likewise, the magnetic susceptibility 
$\chi\sim n_{c}$, as the carriers are activated from the singlet spin-paired ground-state configuration with $\chi\approx 0$.
However, it is important to note that those carriers are coupled antiferromagnetically, as the 
susceptibility exhibits the Curie-Weiss behavior (cf. Fig. 3) with a sizable paramagnetic Curie temperature
$\Theta\simeq -124$~K, and the value of the Curie constant $C=0.5$~emu~K/mol, the value of which is about a half of that 
for the free Ce$^{3+}$ ion. Theoretically, up to a half of the $4f$ moment compensation for the Kondo lattice
comes from the carriers [9].
So, the Kondo gap has a magnetic origin. Second, the system remains Kondo insulating
even by doping CeRhSb with Sn. This means that the interpretation of the Kondo
insulating state as a hybridized-band insulator is not quite accurate, since it represents a {\em collective nonmagnetic
Kondo-lattice\/} state, which cannot formed if the number of valence electrons is diminished below a critical value. 
However, the situation is not easy because of the atomic disorder introduced with the substitution,
the role of which has also been discussed in these low carrier-density systems [12]. Nevertheless, 
the role of disorder does not seem to be a decisive factor in the formation of the Kondo insulating
state, since the gap drops with the increasing $x$. The residual value of the gap
$\Delta\simeq 0.1$~K for the samples with $x=0.10$ and $0.12$ may be due to the disorder. This 
factor may, however, broaden the critical regime to certain interval around $x\sim 0.12$, as is seen in 
experiment, since the critical behavior of $\chi$ is seen even for $x=0.16$ (cf. Fig. 5).

\begin{figure}
\includegraphics[width=0.4\textwidth]{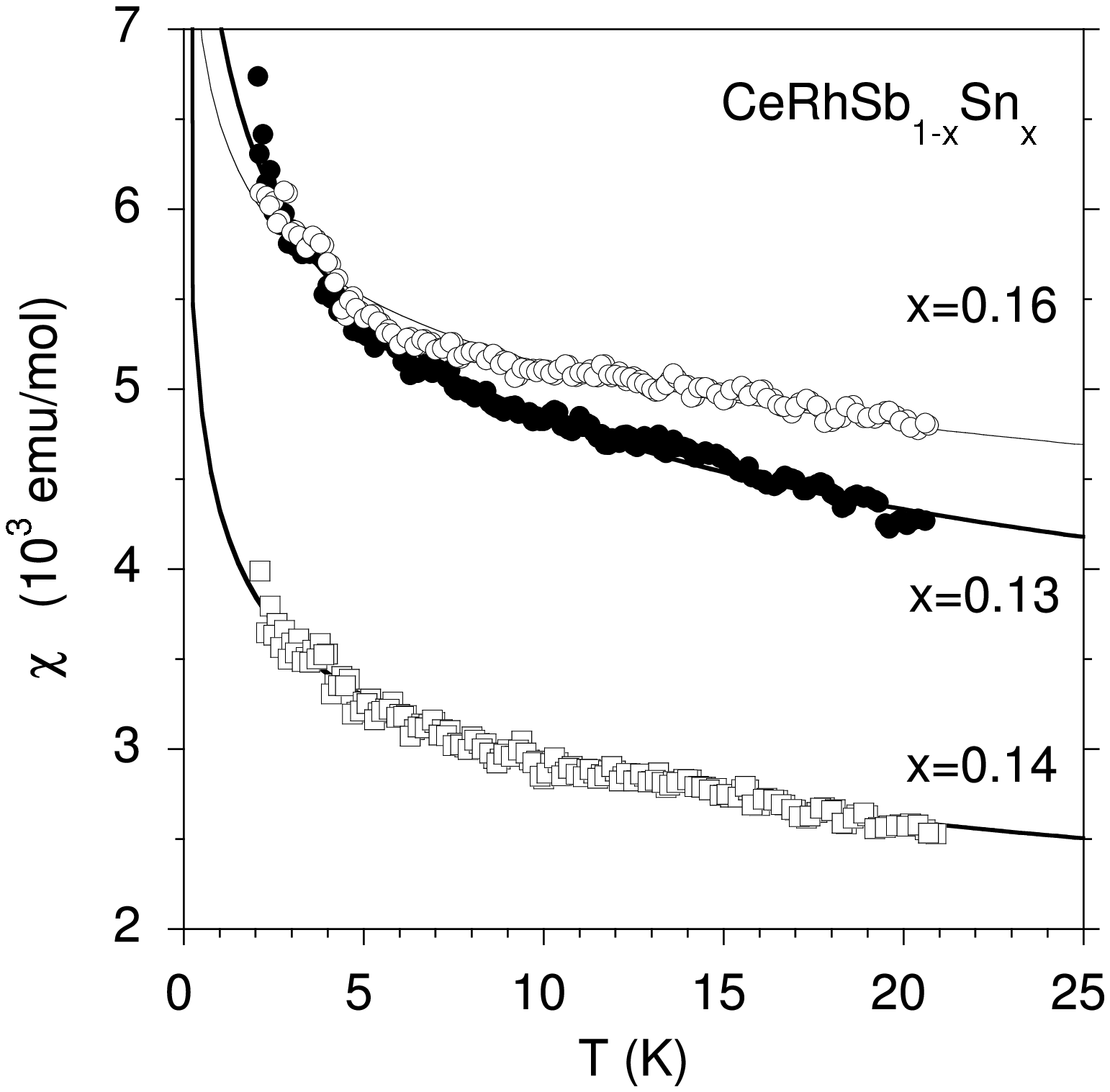}% Here is how to import EPS art
\caption{\label{fig:r5} Magnetic susceptibility for $T< 10$~K and the fitted scaling $\sim T^{-m}$ (see main text).}
\end{figure}

In conclusion, we have shown that the Kondo semiconductors CeRhSb$_{1-x}$Sn$_{x}$, with $x\leq 0.12$ 
are characterized not only by the gap in the conductivity and the vanishing paramagnetic susceptibility
for $T\rightarrow 0$, but also by the universal scaling law 
$\rho\chi=const(x)$. Also, the system undergoes the nonmagnetic Kondo insulator-metal transition at $x\approx 0.12$.
On the metallic side, a novel quantum critical point and NFL behavior has been discovered for the samples $x\rightarrow 0.12+$, 
that is specified by the power-law increase $(T^{-\alpha})$ of the resistivity, the susceptibility, and
the specific heat at lower temperatures. 
The simultaneous divergences of both $\rho$ and $\chi$ at QCP illustrates nicely the circumstance that both
the onset of the KI gap and the magnetic critical fluctuations appearance coexist then.
The transition is markedly different from the Mott-Hubbard transition.
The non-Fermi liquid behavior emerges with $T\rightarrow 0$ from an overall Fermi-liquid state.
Also, the quantum critical behavior described here as a function of the carrier number differs from that 
appearing [13] at the paramagnetic heavy fermion - antiferromagnetic metal boundary.

The support of KBN through Grant No. 2 P03B 050 23 is acknowledged. J.S. thanks the Polish Foundation for Science (FNP)
for a senior fellowship for the years 2003-2006.

%%\newpage %Just because of unusual number of tables stacked at end

\begin{thebibliography}{99}
\bibitem{7}
The drop of resistivity below $T_{max}$ results from a formation of a coherent Anderson-Kondo lattice. The $\rho$ increase $\sim \ln T$ for $T>T_{max}$
is usually attributed to the scattering localizing $4f$ electrons.
\end{thebibliography}
%\newpage

%Fig.~\ref{fig:wide} is a figure that is too wide for a single column,
%so instead the \texttt{figure*} environment has been used.
%\begin{figure*}
%\includegraphics{Fig2.ps}% Here is how to import EPS art
%\caption{\label{fig:wide}Use the figure* environment to get a wide
%figure that spans the page in \texttt{twocolumn} formatting.}
%\end{figure*}

\end{document}